# Temperature dependence of spin diffusion length and spin Hall angle in Au and Pt

Miren Isasa[1], Estitxu Villamor[1], Luis E. Hueso[1,2], Martin Gradhand[3] and Fèlix Casanova[1,2]

[1]CIC nanoGUNE, 20018 Donostia-San Sebastian, Basque Country, Spain.
[2]IKERBASQUE, Basque Foundation for Science, 48011 Bilbao, Basque Country, Spain.
[3]H. H. Wills Physics Laboratory, University of Bristol, Bristol BS8 1TL, United Kingdom

## Abstract

We have studied the spin transport and the spin Hall effect as a function of temperature for platinum (Pt) and gold (Au) in lateral spin valve structures. First, by using the spin absorption technique, we extract the spin diffusion length of Pt and Au. Secondly, using the same devices, we have measured the spin Hall conductivity and analyzed its evolution with temperature to identify the dominant scattering mechanisms behind the spin Hall effect. This analysis confirms that the intrinsic mechanism dominates in Pt whereas extrinsic effects are more relevant in Au. Moreover, we identify and quantify the phonon-induced skew scattering. We show that this contribution to skew scattering becomes relevant in metals such as Au, with a low residual resistivity.

## I. Introduction

*Spintronics* is a rapidly growing research area that aims at using and manipulating not only the charge, but also the spin of the electron. Sophisticated applications such as hard disk read heads and magnetic random access memories have been introduced in the last two decades. A new generation of devices could be achieved with pure spin currents, which are an essential ingredient in an envisioned spin-only circuit that would integrate logics and memory [1]. Therefore, it is of utmost importance to create, transport and detect pure spin currents. Despite several approaches for the generation of spin currents, electrical spin injection is preferred for the integration of spintronic devices into electronics, leading to ferromagnetic materials being the most widely used source of spin currents [2,3,4,5,6,7,8]. Currently, another promising spin-dependent phenomenon is being studied for the spin current generation: the spin Hall effect (SHE). Even if the SHE was predicted theoretically by Dyakonov and Perel more than 40 years ago [9] and revisited by Hirsch more than a decade ago [10], it took a bit longer to observe the first direct experimental evidences in metals [11,12,13]. The SHE is the equivalent to the anomalous Hall effect (AHE) in ferromagnets, but in a nonmagnetic material. When an unpolarized charge current flows in a nonmagnetic conductor, the spin-up and spin-down electrons are deflected in opposite directions due to spin-orbit coupling (SOC). This deflection causes a spin accumulation at the edges of the metal, resulting in a pure spin current in the transverse direction to the charge current (SHE). The reciprocal effect, known as the inverse SHE (ISHE), refers to the transverse charge current created from the flow of a pure spin current. The efficiency of a metal to convert charge current into spin current and vice versa is characterized by the spin Hall angle.

The origin of the SHE has been attributed to three different contributions [14]: i) intrinsic, in which the SOC, proportional to $Z^4$ where Z is the atomic number, is inherent to the electronic structure of the material; ii) skew scattering, an extrinsic mechanism where spin-dependent scattering arises due to the effective SOC of impurities in the lattice; and iii) side jump, also extrinsic and only observed at high impurity concentrations [15]. Despite the extensive theoretical debate on the magnitude of the individual mechanisms in different metals [16,17,18], accompanied by experimental work [19,20,21], the quantitative role of each contribution for any specific system often remains unclear. Nevertheless, the interest into the SHE is clear: spin currents can be generated and detected without using either FM electrodes nor applying an external magnetic field, resulting in a great technological advantage [22,23]. Understanding the underlying physics of the effect to search for materials that provide a large effect has thus become an important topic in spintronics.

In this work, we study the spin transport and the SHE in two different transition metals (TMs). One is platinum (Pt); even though it is one of the prototype metals to exploit the SHE [11,13,19,21], there is still a large controversy regarding the magnitude of the spin Hall angle [24]. The other is gold (Au), which is interesting because very contradicting spin Hall angle values have been reported [20,25,26,27,28]. In addition, Au shows a relatively large spin diffusion length in spite of a strong SOC [3,5,6,26]. The use of lateral spin valve (LSV) devices in which the spin current is created by electrical spin injection from a FM electrode, transported through a non-magnetic (NM) channel and absorbed into a TM wire allows us to obtain both the spin diffusion length (via the spin absorption) and the spin Hall angle (via the ISHE) of the TM [19,21,26,29,30]. Moreover, we measure and analyze the temperature dependence of the spin Hall angle in order to separate the different contributions to the SHE for Pt and Au. Whereas intrinsic mechanisms dominate in Pt, extrinsic effects are more relevant in Au. Most importantly, the low residual resistivity of Au allows the detection of the phonon contribution to the skew scattering. Our careful analysis enables the quantification of this contribution.

## II. Experimental details

We fabricated our devices by multiple-step e-beam lithography on top of a $SiO_2$ (150 nm)/Si substrate, followed by metal deposition and lift-off. These devices consist of two copper (Cu)/permalloy (Py) LSVs, each one with the same separation in between the Py electrodes ($L$~630 nm), one of them having a TM wire in between the electrodes (see Fig. 1(a)). In the first lithography step, the two pairs of FM electrodes were patterned with different widths, ~110 nm and ~160 nm, in order to obtain different switching magnetic fields and 35 nm of Py were e-beam evaporated. In the second lithography step, the middle wire in between the electrodes was patterned and Pt or Au was deposited. The 15-nm-thick and ~150-nm-wide Pt wire was deposited by magnetron sputtering, whereas the 80-nm-thick and ~140-nm-wide Au wire was grown by e-beam evaporation at a base pressure of $\leq 1 \times 10^{-8}$ mbar. In this case, a 1.5-nm-thick Ti layer was deposited before Au in order to avoid adhesion problems. In the third lithography step, a ~150-nm-wide NM channel was patterned and Cu was thermally evaporated with a base pressure of $\leq 1 \times 10^{-8}$ mbar. Different Cu thicknesses of 60, 100 and 145 nm were used in the devices. Before the Cu deposition, the Py and TM wire surfaces were cleaned by Ar-ion milling to ensure transparent contacts.

All non-local transport measurements described in the following have been carried out in a liquid-He cryostat (applying an external magnetic field $H$ and varying the temperature) using a "DC reversal" technique [31].

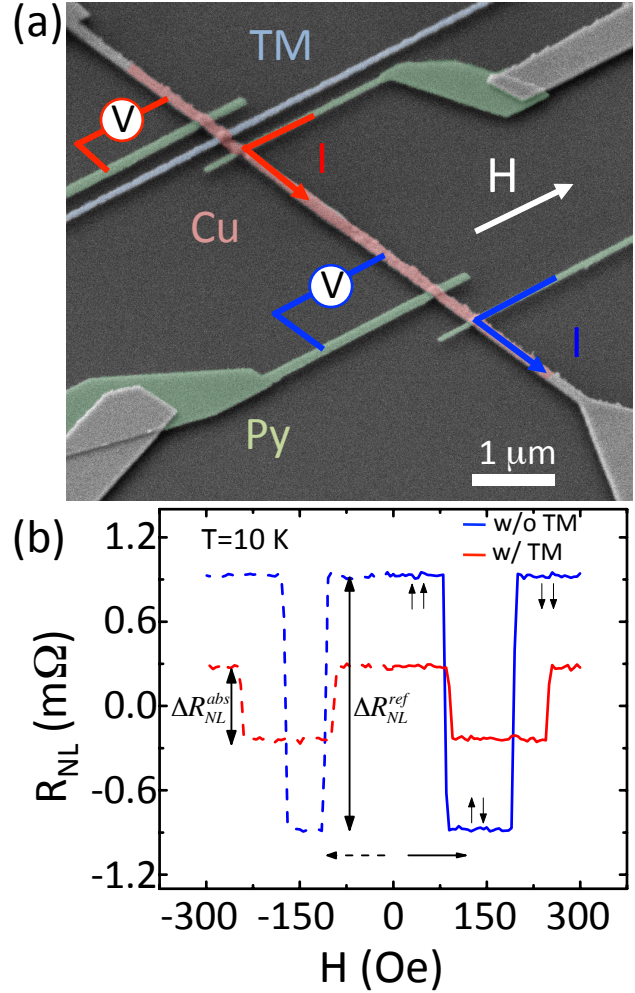

FIG. 1. (Color online) (a) Colored SEM image of two Py/Cu LSVs, one of them with a TM wire in between the Py electrodes and the other one without. The measurements configuration, the direction of the applied magnetic field ($H$) and the materials (Py, Cu and TM) are shown. (b) Nonlocal resistance as a function of $H$ at 10 K for a Py/Cu LSV without (blue line) and with (red line) the TM wire in between the Py electrodes. In this case, TM is Au. The solid (dashed) line represents the increasing (decreasing) sweep of $H$.

## III. Results and discussion

### A. Spin diffusion length

In order to create a pure spin current in a LSV device (Fig. 1a), a nonlocal configuration must be used [2,3,4,5,6,7,8]. When a spin-polarized current is injected from a FM electrode, a spin accumulation is built at the interface between the NM channel and the FM. This spin accumulation diffuses away from the interface, creating a pure spin current which is detected as a voltage by a second FM electrode. From the normalization of the detected voltage $V$ to the injected current $I$, the nonlocal resistance is defined ($R_{NL} = \frac{V}{I}$). This value changes sign when the relative magnetization of the FMs is switched from parallel to antiparallel by sweeping $H$. The change from positive to negative $R_{NL}$ is defined as the spin signal $\Delta R_{NL}$ (Fig. 1b). If a TM wire is placed in between the electrodes, part of the spin current that is diffusing through the NM channel will be absorbed into the TM, thus modifying the detected $\Delta R_{NL}$. The spin absorption (SA) technique (Fig. 1a) [19,21,26,29,30] is based on the comparison of $\Delta R_{NL}$ measured in a conventional FM/NM LSV (reference signal, $\Delta R_{NL}^{ref}$) to the $\Delta R_{NL}$ measured in a LSV when a TM wire is placed in between the FM electrodes (absorbed signal, $\Delta R_{NL}^{abs}$). From the one-dimensional spin diffusion model, the ratio $\eta$ between both signals can be calculated as [29]:

$$\eta = \frac{\Delta R_{NL}^{abs}}{\Delta R_{NL}^{ref}} = \frac{R_{TM}\sinh(L/\lambda_{NM}) + R_{TM}\frac{R_{FM}}{R_{NM}}e^{(L/\lambda_{NM})} + \frac{R_{TM}}{2}\left(\frac{R_{FM}}{R_{NM}}\right)^2 e^{(L/\lambda_{NM})}}{R_{NM}(\cosh(L/\lambda_{NM})-1) + R_{TM}\sinh(L/\lambda_{NM}) + R_{FM}\left[e^{(L/\lambda_{NM})}\left(1+\frac{R_{FM}}{2R_{NM}}\right)\left(1+\frac{R_{TM}}{R_{NM}}\right)-1\right]} \quad (1)$$

where $R_{TM} = \frac{\rho_{TM}\lambda_{TM}}{w_{TM}w_{NM}\tanh(t_{TM}/\lambda_{TM})}$, $R_{NM} = \frac{\rho_{NM}\lambda_{NM}}{2w_{NM}t_{NM}}$ and $R_{FM} = \frac{\rho_{FM}\lambda_{FM}}{(1-\alpha_{FM}^2)w_{NM}w_{FM}}$ are the spin resistances, $\lambda_{TM(NM,FM)}$, $\rho_{TM(NM,FM)}$, $w_{TM(NM,FM)}$ and $t_{TM(NM)}$ are the spin diffusion length, resistivity, width and thickness of the TM(NM,FM), $\alpha_{FM}$ is the spin polarization of the FM and $L$ is the separation between the FM electrodes. Since $R_{FM}$ and $R_{NM}$ values are well known from our previous work [7,8], $\lambda_{TM}$ can be obtained from Eq. (1).

For TM=Pt, we measured $\rho_{Pt}$= 25.0 µΩ cm (39.7 µΩ cm) at 10 K (300 K), which gives $\lambda_{Pt}$ = 3.4 ± 0.3 nm (2.0 ± 2.2 nm) (see Fig. 2a and inset). If we compare the $\lambda_{Pt}$ value at low temperatures to the value measured by Morota *et al.* [21] with the same SA technique, we obtain a shorter value, most likely due to the fact that we have a 2.5 times more resistive Pt. The $\lambda_{Pt}$ value at 300 K is comparable to values reported in literature using different techniques (1.2 – 3.7 nm, see Table I).

Table I. Spin diffusion length and spin Hall angle for Pt and Au extracted from the literature and this work using different methods (lateral spin valve=LSV, spin pumping=SP, spin-torque ferromagnetic resonance=ST-FMR, Hall Cross=HC and spin absorption=SA). Temperature and resistivities are included. * The value reported in the original paper is twice this value due to a factor of 2 difference in the $\theta_{SH}$ definition.

| Material | T (K) | ρ (µΩ cm) | λ (nm) | θ (%) | Method | Ref. |
|---|---|---|---|---|---|---|
| Pt | 300 | 39.7 | 2.0±2.2 | 1.0±1.8 | SA | This work |
|  | 300 | - | 1.4 | 9±2 | SP | 27 |
|  | 300 | 25 | 1.2 | 8.6±0.5 | SP | 33 |
|  | 300 | 20 | 1.4±0.3 | > 5 | ST-FMR | 24 |
|  | 300 | 41.3 | 3.7±0.2 | 4* | SP | 34 |
|  | 300 | 17.3±0.6 | 3.4±0.4 | 5.6±1.0 | SP | 35 |
|  | 10 | 25.0 | 3.4±0.3 | 0.9±0.2 | SA | This work |
|  | 10 | 10 | 10±2 | 2.4±0.6 | SA | 21 |
|  | 8 | - | 1.6 | - | SP | 33 |
| Au | 300 | 8.07 | 32±5 | <0.04 | SA | This work |
| | 300 | 5 | 35 | 0.25±0.1 | SP | 25 |
| | 300 | - | 35 | 0.8±0.1 | SP | 27 |
| | 295 | 3.89 | 36 | <0.27 | HC | 20 |
| | 77 | 3.5 | 98 | - | LSV | 6 |
| | 15 | 4 | 85 | - | LSV | 5 |
| | 10 | 3.62 | 53±2 | 0.21±0.07 | SA | This work |
| | 10 | - | 63±15 | - | LSV | 3 |
| | 10 | 4.0 | 33±9 | 1.0±0.2 | SA | 26 |
| | 4.5 | 2.07 | 65 | < 0.23 | HC | 20 |

For TM=Au, we measured $\rho_{Au}$= 3.62 μΩ cm (8.07 μΩ cm) at 10 K (300 K), plotted in the inset of Fig. 2b. In this case, however, we have to correct the definition for $R_{TM}$ as in the definition described above we are assuming $w_{NM} \gg \lambda_{TM}$ [29], but from literature values we expect $w_{NM} \sim \lambda_{NM}$, in the particular case of Au (see table I). From the general definition of the spin resistance, $R_s = \frac{\rho \lambda^2}{V}$ where $V$ corresponds to the volume in which the spin current diffuses [32], we derive $R_{Au} \approx \frac{\rho_{Au} \lambda_{Au}^2}{w_{Au} t_{Au}(w_{NM}+2\lambda_{Au})}$. Using this definition for $R_{Au}$ in Eq. (1) we obtain $\lambda_{Au}$ = 53 ± 2 nm (32 ± 5 nm) at 10 K (300 K), as plotted in Fig. 2b. These values are in good agreement with those reported in literature (see table I).

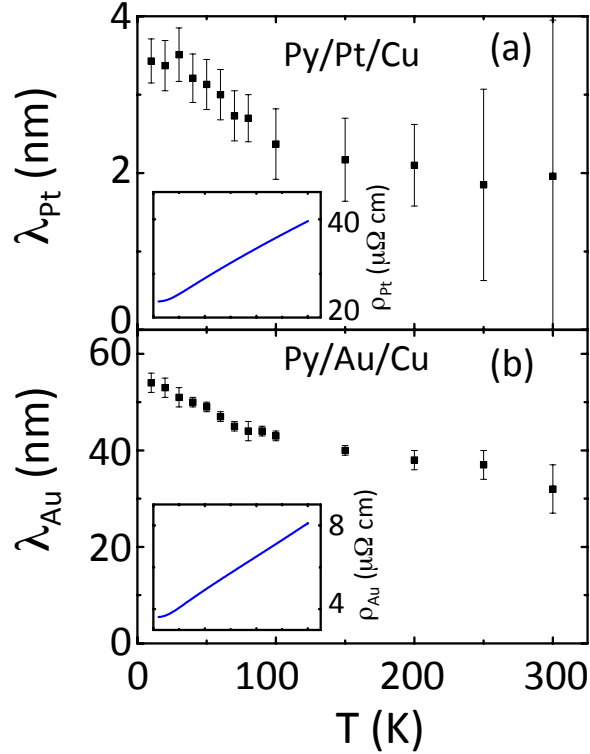

FIG. 2. Spin diffusion length of (a) Pt and (b) Au as a function of temperature obtained from the spin absorption experiment. Insets: (a) Pt and (b) Au resistivity as a function of temperature. Note that the temperature scale in the inset is the same as the temperature scale in the main figure.

Obtaining an accurate value of $\lambda_{TM}$ is a matter of utmost importance to determine the correct magnitude of the SHE, as will be evidenced in the following section.

## B.  Spin Hall angle

The ISHE was measured in Pt and Au using the same devices in which the spin diffusion length was obtained, but changing the measurement configuration as indicated in Fig. 3a. When we inject a current $I_c$ from the Py electrode, the spin accumulation built at the Py/Cu interface diffuses away creating a pure spin current. Part of the spin current that propagates along the Cu is absorbed perpendicularly into the TM wire, resulting in a measurable voltage due to the ISHE [13,19,21,26,29,30]. Note that now the spin polarization of the spin current is parallel to the hard axis of the Py electrode. When a magnetic field is applied along that direction, the measured resistance exhibits a linear increase with increasing the applied field and it saturates above the saturation field of the Py (Fig. 3b). This saturation field can be separately confirmed from the anisotropic magnetoresistance (AMR) measured on the same Py electrode (Fig. 3c). The change in resistance between the two saturated regions at large negative and positive $H$ is twice the inverse spin Hall signal ($2\Delta R_{ISHE}$). Figure 3b shows that $\Delta R_{ISHE}$ is much larger for Pt than for Au, although the sign is the same for both. The spin Hall conductivity $\sigma_{SH}$ is the spin current response to an electric field and, for our device geometry, can be calculated as [21]:

$$\sigma_{SH} = \sigma_{TM}^2 \frac{w_{TM}}{x_{TM}} \left(\frac{I_c}{\overline{I_s}}\right) \Delta R_{ISHE} \qquad (2)$$

where $\sigma_{TM}$ is the charge conductivity of the TM and $x_{TM}$ is a correction factor that takes into account the current that is shunted through the Cu, due to the lower resistivity of Cu compared to the resistivity of the TM wire. $x_{TM}$ is obtained from a different measurement in which the resistance of a TM wire is measured with and without a Cu wire in between the voltage probes [30]. For the case of Au, we obtain $x_{Au} = 0.81$ (0.46), while for Pt we get $x_{Pt} = 0.30$ (0.25), at 10 K (300 K). $\overline{I_s}$ is the effective spin current that contributes to the ISHE and is given by [29]:

$$\frac{\overline{I_s}}{I_c} = \frac{\lambda_{TM}}{t_{TM}} \frac{(1-e^{-t_{TM}/\lambda_{TM}})^2}{1-e^{-2t_{TM}/\lambda_{TM}}} \frac{\alpha_{FM} R_{FM}\left[\sinh(d/2\lambda_{NM}) + \frac{R_{FM}}{2R_{NM}}e^{(d/2\lambda_{NM})}\right]}{R_{NM}[\cosh(d/\lambda_{NM})-1] + R_{FM}\left[e^{d/\lambda_{NM}}\left(1+\frac{R_{FM}}{2R_{NM}}\right)\left(1+\frac{R_{TM}}{R_{NM}}\right)-1\right] + R_{TM}\sinh(d/\lambda_{NM})} \qquad (3)$$

where $d$ is the distance between the Py electrode and the TM wire.

From $\sigma_{SH}$, the spin Hall resistivity is defined as $\rho_{SH} = -\sigma_{SH}/(\sigma_{TM}^2 + \sigma_{SH}^2)$. Assuming $\rho_{TM} \gg \rho_{SH}$, we can approximate it to $\rho_{SH} \approx -\sigma_{SH}/\sigma_{TM}^2$. The spin Hall angle, $\theta_{SH}$, which quantifies the magnitude of the SHE, can be written in terms of either $\sigma_{SH}$ or $\rho_{SH}$: $\theta_{SH} = \frac{\sigma_{SH}}{\sigma_{TM}} = \frac{-\rho_{SH}}{\rho_{TM}}$. As can be deduced from Eqs. 2 and 3, an incorrect value of $\lambda_{TM}$ would strongly affect the obtained value of $\sigma_{SH}$ and $\theta_{SH}$, an issue widely discussed before [24].

For the case of TM=Pt, two different LSV devices have been fabricated, one with $t_{Cu}$ = 60 nm and $d$ = 280 nm and the other with $t_{Cu}$ = 100 nm and $d$ = 310 nm. As shown in Figs. 4a and 4b, the geometrical parameters do not affect the obtained $\sigma_{SH}$ and $\theta_{SH}$ values as a function of temperature, demonstrating consistent results with different devices. From the measurements at 10 K, we obtain $\theta_{SH} \approx 0.9 \pm 0.2\%$ in reasonable agreement with values reported using the same technique [21]. When increasing the temperature, $\sigma_{SH}$ is constant, whereas $\theta_{SH}$ increases monotonically up to $\theta_{SH} \approx 1.0 \pm 1.8\%$ at 300 K. At this temperature, only $\theta_{SH}$ values determined by other techniques have been reported, which are substantially larger (between 4 and 9%, see Table I). This discrepancy between different techniques estimating the spin Hall angle has been discussed before [24] and no final conclusion has been reached.

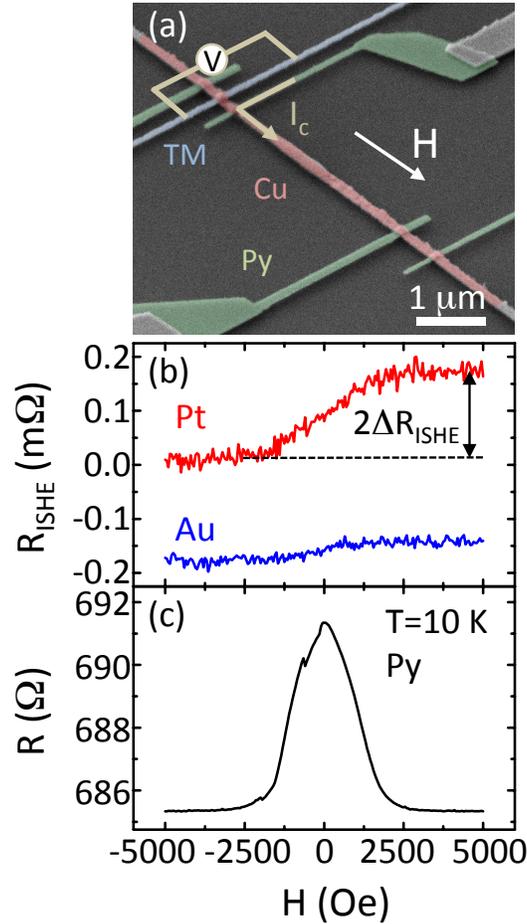

FIG 3 (a) Colored SEM image of the same device shown in Fig. 1a used now to measure the ISHE. The materials (Py, Cu and TM), the direction of the magnetic field ($H$) and the measurement configuration for ISHE are shown. (b) Non-local resistance for Pt (red line) and Au (blue line) as a function of $H$ measured at 10 K in the ISHE configuration shown in (a). (c) Resistance as a function of $H$ (applied as shown in (a)) for the Py electrode used for spin injection, measured at 10 K.

For the case of TM=Au, we choose a 145-nm-thick Cu channel and two different distances ($d$ = 180 nm and $d$ = 260 nm) between the Py electrode and Au wire. As plotted in Figs. 4c and 4d, reproducible $\sigma_{SH}$ and $\theta_{SH}$ values as a function of temperature are obtained when varying $d$, showing consistent results with different devices. From measurements at 10 K, we obtain $\theta_{SH} \approx 0.21 \pm 0.07$ %. When increasing the temperature, both $\sigma_{SH}$ and $\theta_{SH}$ decrease strongly and go below the measurable threshold for $T > 200$ K. This temperature dependence is similar to what is reported in Ref. 26, but with slightly lower values in our case. We thus expect $\theta_{SH} < 0.04$ % at 300 K. Again, this value clearly differs from results obtained with the spin pumping technique, in which values between 0.25 and 0.8% at room temperature are reported (see Table I).

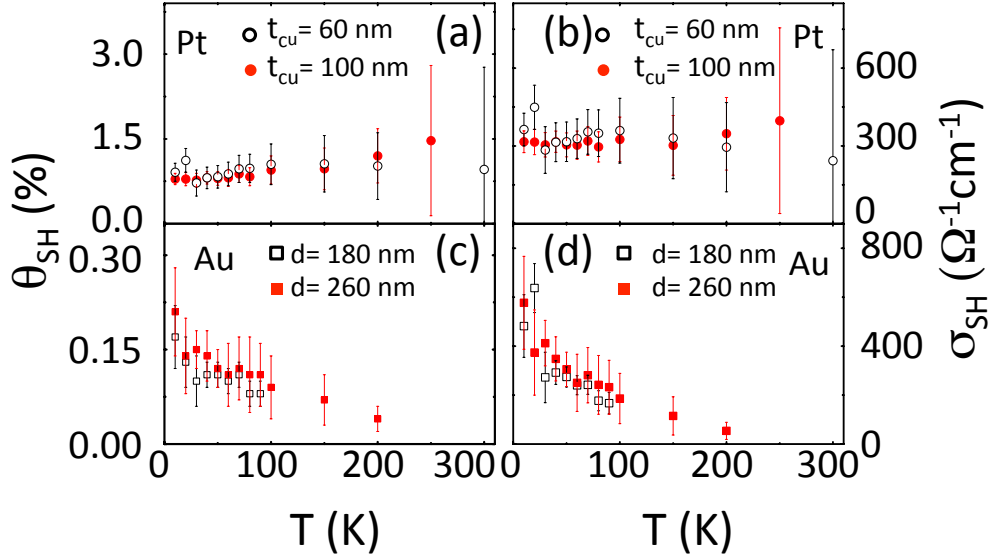

FIG 4. Spin Hall angle (a) and spin Hall conductivity (b) of Pt as a function of temperature obtained from two devices with $t_{Pt}=15$ nm and different $t_{Cu}$ (see legend). Spin Hall angle (c) and spin Hall conductivity (d) of Au as a function of temperature obtained from two devices with $t_{Au}=80$ nm and different $d$ (see legend).

In order to gain a deeper understanding of the mechanisms behind the SHE, we look into its temperature dependence. Whereas the intrinsic mechanism is related to the band structure of the metal, extrinsic mechanisms could include skew scattering and side jump [14]. Up to now, the intrinsic mechanism has been reported to dominate over extrinsic mechanisms in 4d and 5d transition metals, such as Nb, Ta, Mo, Pd and Pt [16,21]. In our metallic systems, with low impurity concentrations, the skew scattering mechanism dominates over side jump [15,37]. Therefore, only skew scattering will be taken into account as extrinsic contribution. In analogy to the AHE, the total spin Hall conductivity is calculated by considering the intrinsic and extrinsic contribution as parallel channels ($\sigma_{SH} = \sigma_{SH}^{int} + \sigma_{SH}^{ext}$) and the various extrinsic scattering mechanisms, impurities and phonons, as independent scattering sources forming a serial resistor circuit ($\rho_{SH}^{ext} = \rho_{SH}^{imp} + \rho_{SH}^{phon}$) [38,39]. This leads us to:

$$\sigma_{SH} = \sigma_{SH}^{int} + \sigma_{SH}^{ext} = \sigma_{SH}^{int} - \frac{\rho_{SH}^{imp} + \rho_{SH}^{phon}}{\left(\rho_{TM}^{imp} + \rho_{TM}^{phon}\right)^2 + \left(\rho_{SH}^{imp} + \rho_{SH}^{phon}\right)^2} \quad (4)$$

where $\rho_{TM}^{imp}$ and $\rho_{TM}^{phon}$ are the impurity and phonon contributions to the total resistivity, respectively ($\rho_{TM} = \rho_{TM}^{phon} + \rho_{TM}^{imp}$). Taking into account that $\rho_{SH} \ll \rho_{TM}$, we can rewrite Eq. (4) as:

$$\sigma_{SH} = \sigma_{SH}^{int} - \frac{\rho_{SH}^{imp} + \rho_{SH}^{phon}}{\rho_{TM}^2} \quad (5)$$

In the case that the intrinsic term dominates ($\sigma_{SH}^{int} \gg \frac{\rho_{SH}}{\rho_{TM}^2}$), $\sigma_{SH}$ is independent from the mean free path for scattering and $\theta_{SH}$ depends on $\rho_{TM}$ in the form of $\theta_{SH} \propto \rho_{TM}$. Therefore, $\sigma_{SH}$ is temperature independent and $\theta_{SH}$ will increase linearly with $T$. This is the behavior that we observe for Pt (Figs. 4a and 4b) confirming that the intrinsic contribution is dominant. However, the decrease of $\sigma_{SH}$ and $\theta_{SH}$ that we observe with $T$ for the case of Au (Figs. 4c and 4d) cannot be explained by a dominating intrinsic contribution. Similar experimental results with a strong temperature dependence of $\theta_{SH}$ in Au have been recently reported by Niimi et al. [26], although the effect is attributed to an intrinsic mechanism.

Realistically, we have to take into account both intrinsic and extrinsic contributions, which we will quantify for Pt and Au. In order extract the individual contributions, we rewrite Eq. (5) in terms of $\rho_{SH}$ assuming, in a first approximation, that phonon skew scattering, $\rho_{SH}^{phon}$, is negligible for the spin Hall resistivity [40]:

$$-\rho_{SH} = \sigma_{SH}^{int} \rho_{TM}^2 - \rho_{SH}^{imp} \quad (6)$$

If we plot $-\rho_{SH}$ against $\rho_{TM}^2$, we can directly fit a linear function in which the slope gives the magnitude of the intrinsic contribution and the onset the extrinsic one (Figs. 5a and 5b). The values that we extract from this fitting are summarized in Table II, where the relation $\sigma_{SH}^{imp} \approx -\rho_{SH}^{imp}/(\rho_{TM}^{imp})^2$ has been used.

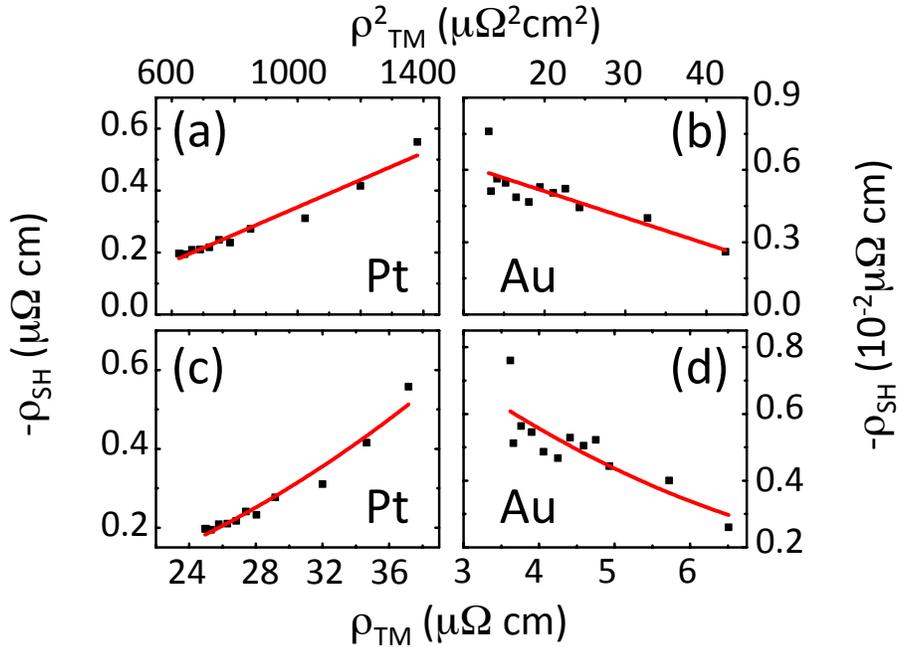

FIG. 5. Spin Hall resistivity as a function of the square of the total resistivity for (a) Pt and (b) Au (black dots). The red solid line is a fit of the data to Eq. (6), where phonon skew scattering contribution is neglected. Spin Hall resistivity as a function of the total resistivity for (c) Pt and (d) Au (black dots). The red solid line is a fit of the data to Eq. (8), taking into account phonon skew scattering contribution.

TABLE II. Summary of the fitting parameters obtained from data plotted in Fig. 5.

|  |  | $\sigma_{SH}^{int}$ $\Omega^{-1}cm^{-1}$ | $\sigma_{SH}^{imp}$ $\Omega^{-1}cm^{-1}$ | $\theta_{SH}^{phon}$ (%) |
|---|---|---|---|---|
| **without phonon** | Pt | 439 ± 29 | 149 ± 40 | - |
|  | Au | -109 ± 24 | 557 ± 41 | - |
| **with phonon** | Pt | 475 | 182 ± 15 | -0.24 ± 0.17 |
|  | Au | 360 | 118 ± 24 | -0.20 ± 0.03 |

As can be seen from Table II, the intrinsic contribution in Pt dominates over the extrinsic one, as expected both from theoretical [16,17] and other experimental work [21], with a magnitude in close agreement with tight-binding calculations (475 $\Omega^{-1}$cm$^{-1}$) from Ref. [16]. On the other hand, the extrinsic contribution in Au dominates over the intrinsic spin Hall conductivity, which is consistent with previous theoretical work [41]. However, we obtain the opposite sign of $\sigma_{SH}^{int}$ for the case of Au compared to Pt, in disagreement with first-principles calculations [17,42,43]. Furthermore, both transition metals have more than half-filled d-bands, pointing to a positive intrinsic spin Hall conductivity as discussed previously [44]. The origin of this unexpected sign is that the temperature dependence that enters in Eq. (5) through $\rho_{TM}$ is thus not enough to account for the strong temperature decay in $\sigma_{SH}$ for Au (Fig. 4d). A possible explanation could be that neglecting the phonon contribution to skew scattering is not a valid simplification. We can thus reintroduce this term, so that Eq. (6) is now

$$-\rho_{SH} = \sigma_{SH}^{int}\rho_{TM}^2 - \rho_{SH}^{Phon} - \rho_{SH}^{imp} \qquad (7)$$

Assuming that skew scattering at phonons $\left(\rho_{SH}^{Phon} \propto \rho_{TM}^{Phon}\right)$ has the same scaling as the skew scattering at impurities $\left(\rho_{SH}^{imp} \propto \rho_{TM}^{imp}\right)$ we can rewrite Eq. (7) as

$$-\rho_{SH} = \sigma_{SH}^{int}\rho_{TM}^2 + \theta_{SH}^{Phon}\left(\rho_{TM} - \rho_{TM}^{imp}\right) + \sigma_{SH}^{imp}\left(\rho_{TM}^{imp}\right)^2 \qquad (8)$$

where $\theta_{SH}^{phon}$ is the phonon contribution to the spin Hall angle, which is temperature independent. By fitting our experimental data to Eq. (8) and fixing the intrinsic spin Hall conductivities from values obtained by tight-binding calculations [16], see Figs. 5c and 5d, we obtain the values reported in Table II. For Au we find a non-zero $\theta_{SH}^{phon}$ value, suggesting that phonon skew scattering might be an important contribution that has to be taken into account. However, a phonon contribution has not been identified up to now, either by studying the SHE in Pt [21], or in analyzing the AHE in Fe [40]. Indeed, the $\theta_{SH}^{phon}$ value we obtain for Pt is compatible with the value obtained for Au, although its contribution is irrelevant and hardly changes the weight of the other contributions (see table II). This observation evidences that the phonon term is not detectable experimentally in Pt. However, for the case of Au, it is clear that adding the phonon contribution involves a substantial change in the rest of the parameters (see table II). One reason to observe it so unambiguously in Au is the low resistivity of this metal. From Eq. (8), it can be clearly seen that the different contributions scale differently with the resistivity. The intrinsic term scales with $\propto \rho_{TM}^2$, so that, in metals with large resistivity, this term will dominate over the rest. The phonon contribution term scales with $\propto (\rho_{TM} - \rho_{TM}^{imp})$, which means that, for small residual resistivities $\rho_{TM}^{imp}$ like in the case of Au, this second term is comparable or higher than the intrinsic term and, therefore, it cannot be disregarded. Finally, the impurity contribution scales with $\propto (\rho_{TM}^{imp})^2$, dominating over the phonon term in metals with higher residual resistivity.

## IV. Conclusions

In summary, we used the spin absorption technique to determine the particularly short spin diffusion length of metals with strong SOC, impossible to extract using conventional LSVs. Additionally, using the same device, we obtained the spin Hall angle for Au and Pt. We find systematically smaller spin Hall angles in comparison to those estimated by the spin pumping and spin-torque ferromagnetic resonance techniques. Moreover, we measured the temperature dependence of the SHE in Pt and Au to study the different contributing mechanisms. Whereas the intrinsic mechanism is the dominant contribution in Pt, for the case of Au extrinsic mechanisms play an important role. In particular, we have reported experimental evidence of a strong decay in the spin Hall angle for Au, which cannot be explained unambiguously by the intrinsic and impurity contributions. Therefore, we show that the phonon skew scattering contribution has to be taken into account as a source for the SHE, especially in materials, such as Au, where the residual resistivity is low. Additional work would be needed to better quantify the phonon-induced skew scattering in Au by systematically varying the residual resistivity.

**Acknowledgments**

We thank Dr. Y. Niimi and Prof. Y. Otani for fruitful discussions. This work is supported by the European Union 7th Framework Programme under the Marie Curie Actions (PIRG06-GA-2009-256470-ITAMOSCINOM) and the European Research Council (257654-SPINTROS), by the Spanish MINECO under Project No. MAT2012-37638 and by the Basque Government under Project No. PI2011-1. M. I. and E. V. thank the Basque Government for a PhD fellowship (Grants No. BFI-2011-106 and No. BFI-2010-163).